\documentclass[sigconf]{acmart}

\copyrightyear{2026}
\acmYear{2026}
\acmConference[CHI '26 Workshop]{Engagement in Digital Health Interventions: Open Questions for Research and Design}{April 13--17, 2026}{Barcelona, Spain}
\acmBooktitle{Proceedings of CHI '26 Workshop: Engagement in Digital Health Interventions: Open Questions for Research and Design, April 13--17, 2026, Barcelona, Spain}

\setcopyright{rightsretained}

\begin{document}

\title{Designing for Engagement: How Self-Determination Theory Can Guide Digital Health Design for User Motivation}

\author{Zheyuan Zhang}
\affiliation{%
  \institution{Dyson School of Design Engineering, Imperial College London}
  \city{London}
  \country{United Kingdom}
}
\email{zheyuan.zhang17@imperial.ac.uk}

\author{Rafael A. Calvo}
\affiliation{%
  \institution{Dyson School of Design Engineering, Imperial College London}
  \city{London}
  \country{United Kingdom}
}
\email{r.calvo@imperial.ac.uk}

\begin{abstract}
User engagement is crucial for the efficacy of digital health and mental health interventions, yet existing design strategies for improving engagement remain heterogeneous, context-specific, and insufficiently grounded in motivational theory. In this workshop paper, we propose a theory-grounded, preliminary design framework that draws on Self-Determination Theory (SDT) and its sub-theory, Organismic Integration Theory (OIT), to guide the design of digital health interventions for sustained user engagement. Informed by existing literature and our own empirical data from surveys (N = 438), interviews (N = 31), and co-design workshops (N = 59) with end users, the framework categorises design strategies across the adoption, interface, and task spheres of the user experience, distinguishing between those that primarily support intrinsic motivation and those that foster autonomous forms of extrinsic motivation. We argue that this distinction is critical: strategies commonly grouped under umbrella terms such as ``gamification'' in fact operate through different motivational channels and should be designed and evaluated accordingly. By clarifying these motivational pathways, our framework aims to support researchers and practitioners in designing digital health interventions that not only facilitate initial uptake but also enhance the internalisation of health behaviours for long-term, sustained engagement. We present this framework as a basis for discussion at this workshop, inviting expert feedback and critique to refine it as a contribution to the field.
\end{abstract}

\begin{teaserfigure}
  \includegraphics[width=\textwidth]{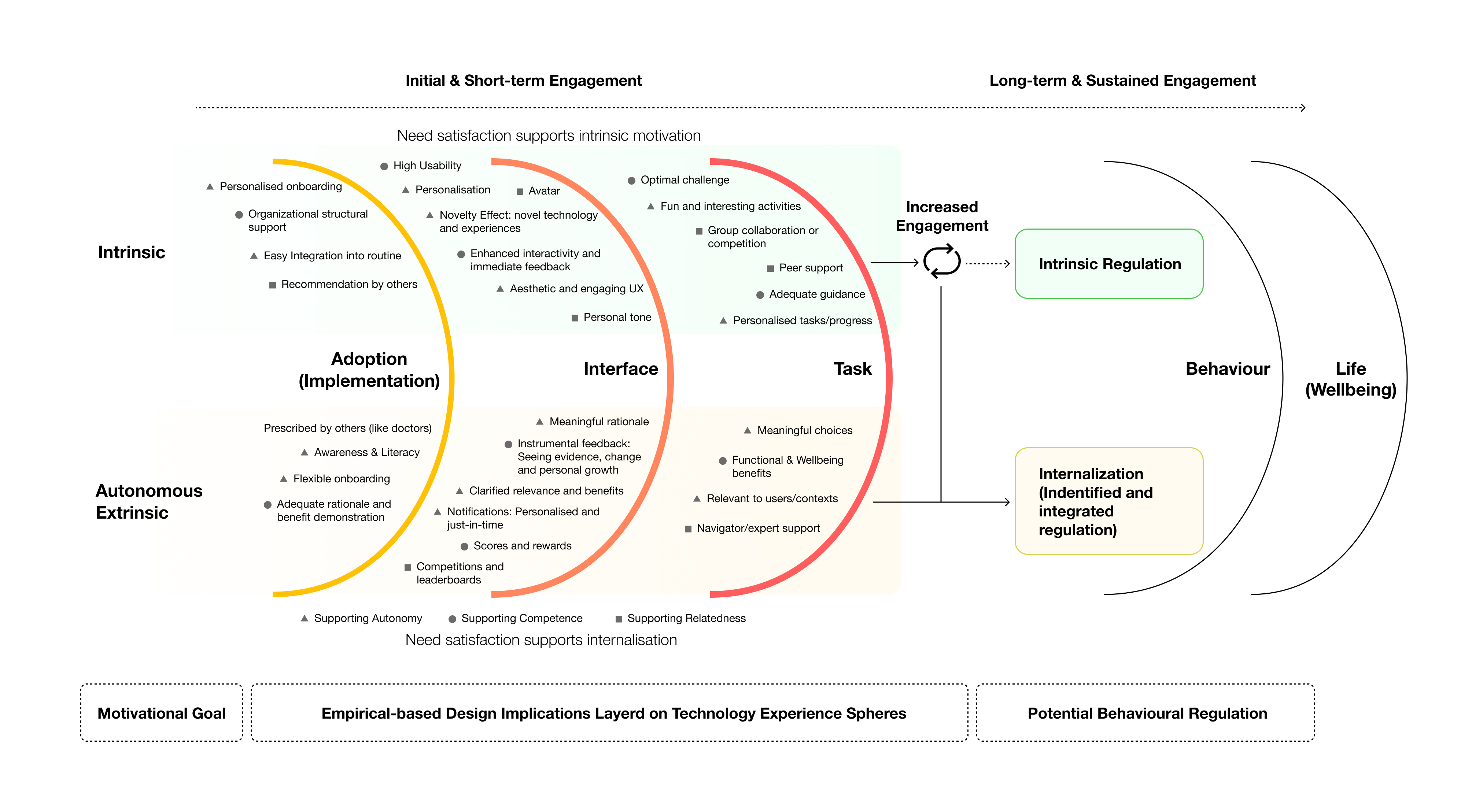}
  \caption{A preliminary framework on how design strategies can drive short- and long-term user engagement in digital health}
  \Description{A framework diagram showing design strategies mapped across METUX experience spheres.}
  \label{fig:framework}
\end{teaserfigure}



\maketitle

\section{Introduction and Background}

User engagement is crucial for digital health and mental health interventions to maximise their efficacy~\cite{borghouts2021barriers}. In the digital health field, studies often investigate and adopt design strategies to improve engagement, such as enhancing usability, providing notifications, improving content relevance, and incorporating personalisation, social, and gamification elements~\cite{gan2022technology,grady2023effectiveness,saleem2021understanding}. Nonetheless, existing strategies and design recommendations remain heterogeneous and context-specific, dispersed across different user populations and intervention technologies~\cite{boucher2024engagement,gan2022technology}. Moreover, relatively little work has empirically linked these strategies to established motivational theory (e.g.,~\cite{alberts2024designing}), limiting the field's ability to provide overarching, theory-grounded design principles for sustained engagement in digital health interventions.

\begin{figure*}[h]
  \centering
  \includegraphics[width=\textwidth]{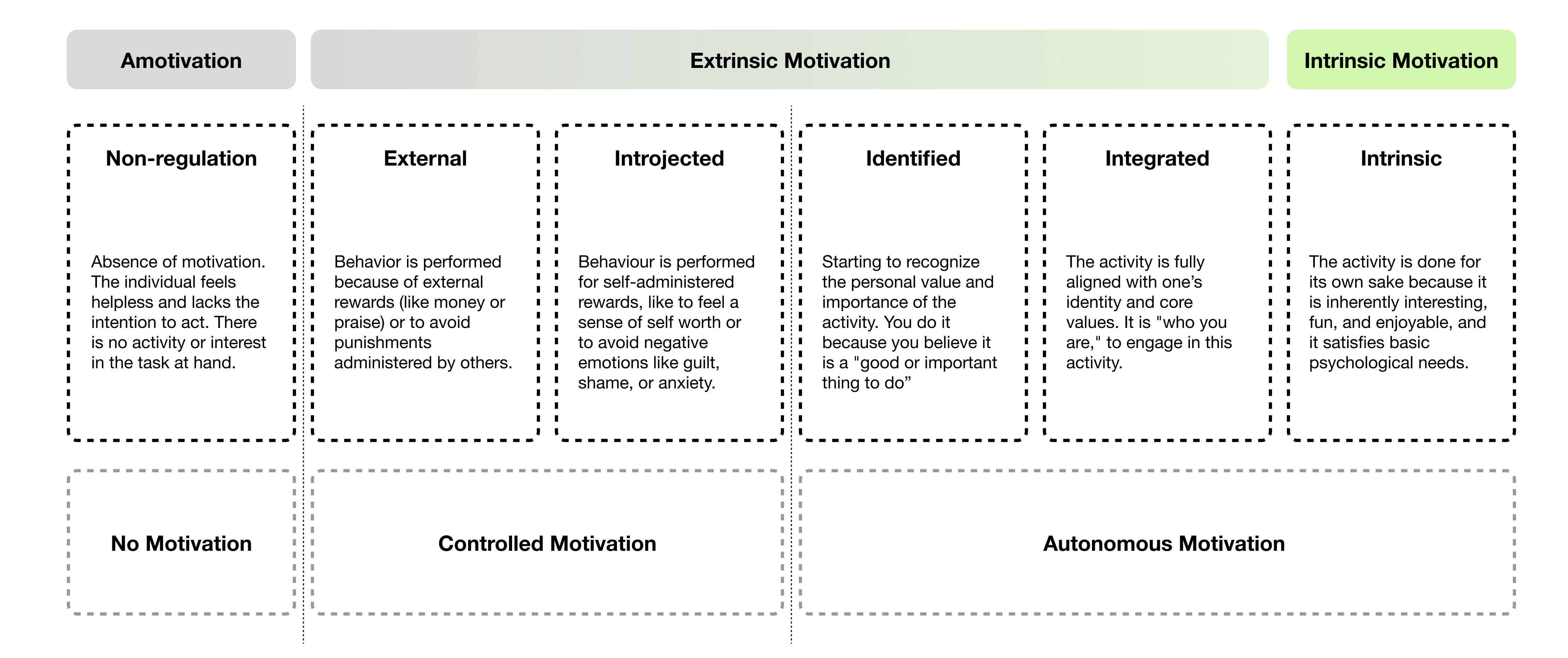}
  \caption{Six different ways motivation to an activity can be regulated, based on SDT and OIT. Adapted from~\cite{ryan2018selfdetermination}}
  \Description{A diagram showing the six types of motivation regulation from SDT and OIT, ranging from amotivation through external, introjected, identified, and integrated regulation to intrinsic motivation.}
  \label{fig:oit}
\end{figure*}

Hence, we propose a theory-grounded framework for designing digital health interventions to promote user engagement. Specifically, we examine how Self-determination Theory (SDT) and its sub-theory, Organismic Integration Theory (OIT), can guide the design of digital health interventions to enhance user engagement. SDT posits that human motivation can be driven by three basic psychological needs: autonomy (a sense of volition and self-endorsement), competence (a sense of effectiveness and mastery), and relatedness (meaningful connectedness with others)~\cite{ryan2023oxford,ryan2000selfdetermination}. OIT, a sub-theory of SDT, posits that an individual's motivation can be regulated in distinct ways depending on their level of autonomy (illustrated in Figure~\ref{fig:oit})~\cite{ryan2018selfdetermination}. Crucially, studies suggest that not only intrinsic motivation but also autonomous extrinsic motivation, including identified and integrated regulation, predict behaviours such as sustained engagement~\cite{bennett2024beyond}. This makes OIT especially applicable to digital health interventions, with unique insights into long-term user engagement~\cite{alberts2024designing}.

\section{Case Study: How Can We Design for Sustained Engagement in Digital Health?}

Based on existing literature on engagement challenges, facilitators, and design strategies~\cite{borghouts2021barriers,carolan2018employees,forbes2023assessing,grady2023effectiveness,jardine2023between,liverpool2020engaging,saleem2021understanding}, as well as our own empirical data collected from surveys (N~=~438) and interviews (N~=~31, reported in~\cite{zhang2025hcp}), and co-design workshops (N~=~59, reported in~\cite{zhang2025beyond} and~\cite{zhang2026understanding}) with end users, we propose a preliminary design framework to categorise design strategies of digital health interventions aimed to enhance users' motivation to engage, as illustrated in Figure~\ref{fig:framework}. We adopted the user experience spheres, such as the Adoption, Interface, Task, Behaviour, and Life, in the METUX model proposed by Peters et al.~\cite{peters2018designing}.

Specifically, we propose that improving engagement with digital health tools requires designing for both intrinsic motivation and autonomous forms of extrinsic motivation. Digital health tools need to support users' basic psychological needs across the adoption/implementation, interface, and task spheres of the product experience. By meeting these needs across each sphere and aligning fun, enjoyable interaction design with users' personally valued health goals, digital health tools can improve the user experience, promote meaningful engagement, and, critically, facilitate the internalisation of target behaviours, enabling long-term, sustained engagement beyond initial uptake.

One aim in proposing this framework is to distinguish design implications that primarily support intrinsic motivation from those that foster autonomous extrinsic motivations. By making this distinction, we clarify that many strategies commonly crowned as improving engagement and retention by existing studies actually serve different motivational functions and should therefore be designed and evaluated with different aims in mind. For example, gamification, or ``gamification elements'', is frequently treated as a general engagement strategy~\cite{rajani2023engagement,venter2022online,zairon2023gamification}. In our interpretation, however, different gamification elements operate through distinct motivational channels. Elements such as avatars, optimal challenge, and immediate (or ``juicy'') feedback mostly support intrinsic motivation by making digital experiences feel enjoyable and fun across multiple experience spheres (as mapped in Figure~\ref{fig:framework}). By contrast, features such as rewards, leaderboards, and competition typically function as extrinsic motivational cues. They can increase motivation to engage, but via a different pathway. And, depending on context and design, may also introduce risks of controlled regulation (e.g., pressure or social comparison).

In this workshop, we hope to discuss this framework with experts and practitioners and gain insights, suggestions, and challenges to this initial proposition. We hope this could be a potential contribution to the digital health field and support future research and industry practices on designing digital health interventions.

\bibliographystyle{ACM-Reference-Format}
\bibliography{references}

\end{document}